\documentclass{article}
\usepackage{amsmath,graphicx,mlspconf,bm,amssymb}

\copyrightnotice{978-1-6654-8547-0/22//\$31.00 {\copyright}2022 IEEE}

\toappear{2022 IEEE International Workshop on Machine Learning for Signal Processing, Aug.\ 22--25, 2022, Xi'an, China}

\title{Over-the-Air Split Learning with MIMO-Based Neural Network and Constellation-Based Activation}
\name{Yuzhi Yang$^*$, Zhaoyang Zhang$^*$, Zhaohui Yang$^{*\dagger}$
\thanks{This work was supported in part by National Key R\&D Program of China under Grant 2020YFB1807101 and 2018YFB1801104, and National Natural Science Foundation of China under Grant U20A20158, and 61725104.}}
\address{$^*$College of Information Science and Electronic Engineering, Zhejiang University, Hangzhou, China\\
Zhejiang Provincial Key Laboratory of Info. Proc., Commun. \& Netw. (IPCAN), Hangzhou, China\\
$^\dagger$Zhejiang Lab, Hangzhou 31121, China.\\
E-mails: \{yuzhi\_yang, ning\_ming, yang\_zhaohui\}@zju.edu.cn}

\begin{document}

\maketitle

\begin{abstract}
This paper investigates a communication-efficient split learning (SL) over  multiple-input multiple-output (MIMO) communication system. 
In particular, we mathematically decompose the inter-layer connection of a neural network (NN) to a series of linear precoding and combining transformations using over-the-air computation (OAC), which synergistically form a linear layer in NNs. 
The precoding and combining matrices are trainable parameters in such a system, whereas the MIMO channel is implicit.
The proposed system eliminates the implicit channel estimation through exploiting the channel reciprocity and properly casting the backpropagation process, significantly saving the system costs and further improving the overall efficiency.
The practical constellation diagrams are used as the activation function to avoid sending arbitrary analog signals as in the traditional OAC system.
Numerical results are illustrated to demonstrate the effectiveness of the proposed scheme.

\end{abstract}
\section{Introduction}
\label{sec:intro}
 \vspace{-.5em}

In future sixth-generation (6G) wireless communication systems, artificial intelligence (AI) will be tightly coupled \cite{6Gchallenges,xu2022edge}, calling for hybrid designs of wireless communication and computation.
Over-the-air computation (OAC) enables computation along with communication, bringing an alternative to the traditional digital communication scheme \cite{OAC}.
In OAC, multiple transmitters simultaneously send local analog signals to a receiver, which obtains the aggregation of received signals through only one single transmission.
Through applying different modulations, OAC can compute the weighted sum or some other large-scale operations, but simple tasks such as geometric mean, polynomial, and Euclidean norm  \cite{OACfunc}.
OAC has been applied to the typical federated learning (FL) scheme \cite{FLOAC}, where OAC is used to calculate the weighted sum of the parameters on different edge devices in FL.

However, FL is a specific learning task, where OAC only acts as an aggregator in the system and does not reach its full potential.
Different from FL, split learning (SL) is a more generalized distributed learning framework, where several edge devices coopertively deploy a neural network (NN).
Due to its flexible framework over wireless networks,  
SL has shown the superiority in AI-empowered systems \cite{review}.
SL distributes a deep NN to many nodes in such systems to ease the computation burden on each piece of equipment, where communication links can be viewed as ideal tubes with certain costs.
SL is also potential in many network for AI applications such as distributed sensing, channel estimation and prediction, location estimation, distributed network optimization, and interference coordination \cite{6Gchallenges}.
In these applications, multiple communication devices cooperate for a particular learning task.
They involve coordination, information exchanges, and distributed algorithms between devices and hence can be viewed as specific SL systems.
To further improve system performance, 
SL systems call for an efficient scheme of joint wireless communication and learning resource design.

To fully utilize the potential of OAC, we propose the multiple-input multiple-output (MIMO)-based NNs and the constellation-based activation functions to implement SL via OAC, coupling SL with the widely applied modules in wireless communications.
A MIMO channel provides a full connection between the input and the output signals and is thus a weighted sum calculator where the channel matrix determines the weight.
On the other hand, the operation of a fully connected layer in a NN is a controllable weighted sum.
To control the MIMO-based OAC system, we can introduce the precoding and the combining procedures into the proposed system.
Both procedures are controllable linear transformations inherent in MIMO wireless communication networks.
As a result, the overall system, including precoding, transmitting, and combining, is equivalent to a controllable linear transformation, i.e., a fully connected layer in NN.

Although the parameters are implicit because of the unknown channel in the proposed system, it can still work well owing to the channel reciprocity.
We observe the mathematical similarity between the channel reciprocity and the forward-backward propagation. 
Through exploiting the relationship in the mathematical forms, we show that the correct gradient can be computed even without accurate channel.
 Moreover, the intermediate results of a NN are uninterpreted even when it works well.  

In traditional OAC, it requires sending arbitrary analog signals, which is hard to produce. 
To solve this problem, the transmitted signal can be modulated by constellation diagram in OAC, which can be viewed as quantifying the original signals. 
Previous works on quantified NNs show that step functions are native activation functions \cite{BNN, DSQ}.
In  \cite{BNN, DSQ}, the authors quantify the sigmoid function or similar activations, which can be equivalent to the multiple-level quantification in forward computation.
We can further extend the step functions to complex numbers and the step function becomes quantification by the nearest point of a quadrature amplitude modulation (QAM) constellation diagram, which is widely applied in existing wireless communication systems.
By applying quantification instead of traditional activation, the system becomes more stable against noise, whereas the performance deteriorates under high-quality channels.

In this paper, we investigate the OAC and constellation-based SL system.
The proposed system synergistically incorporates the MIMO wireless communication system and SL, 
potential in wireless communication-enabled distributed learning systems. 
In the proposed system, the precoding and combining matrices act as the trainable parameters, serving as a fully connected or convolutional layer with the MIMO wireless channel.
The explicit channel estimation is eliminated by exploiting the channel reciprocity, saving the overall system cost.
We also propose constellation-based activation for the proposed system to avoid producing analog signals with arbitrary amplitude in OAC.
Finally, we numerically demonstrate the effectiveness of the proposed system.



\section{System Model}
\subsection{The MIMO OAC-based SL Sysmtem}
\begin{figure}[t]

  \centering
  \centerline{\includegraphics[width=8.5cm]{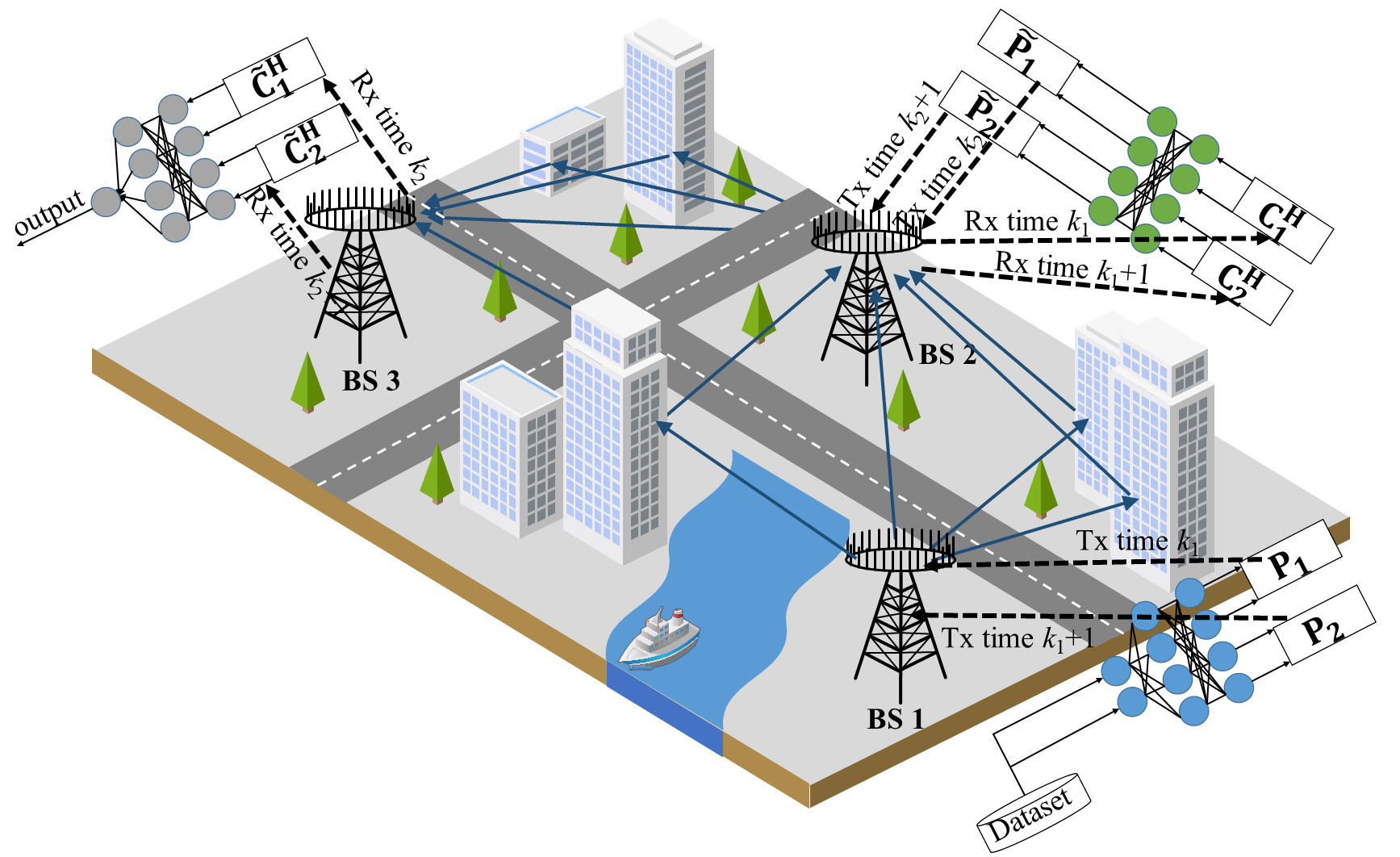}}
%
 \vspace{-.5em}
\caption{The considered MIMO OAC-based SL system.}
 \vspace{-1.5em}
\label{fig::scenario}
\end{figure}
We consider a MIMO OAC-based SL consisting of multiple base stations as shown in Fig. \ref{fig::scenario}.
There are multiple antennas on each base station and the MIMO channels among base stations are quasi-stable.
A NN is segmented and distributed over the base stations.
The base stations process the forward computation of the assigned NN fragment sequentially and conduct the backpropagation in a backward order.

For simplicity, the considered SL system  only has one splitting point unless otherwise stated since a multiple-split system can be decomposed into multiple one-split systems. To simplify the description, we refer to the transmitter and receiver of the forward computation with $N_t$ and $N_r$ antennas as ``transmitter'' and ``receiver'', respectively.
Assume that the forward channel between the transmitter and the receiver is $\mathbf{H}\in\mathcal{C}^{N_r \times N_t}$ with reciprocity, i.e., the backward channel is $\mathbf{H}^T$. Although explicit $\mathbf{H}$ is unknown, $\mathbf{H}$ is assumed to be stable over the transmission of each epoch, and its rank is known to be $r$\footnote{Note that $r$ may be underestimated in practice since some small singular values are considered to be zero. We also note that although the channel is assumed to be stable when designing the system, we numerically demonstrate its efficiency under slowly moving channels in Section \ref{sec::num}.}, which restricts the maximum number of data streams in each transmission.

\subsection{System Implementation}
\begin{figure*}[t]
  \centering
  \centerline{\includegraphics[width=15cm]{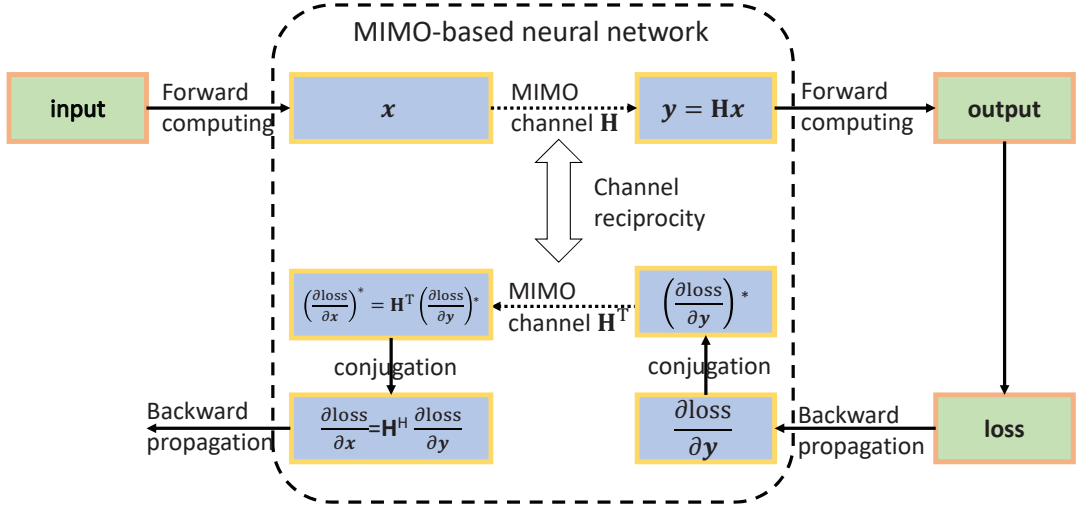}}
 \vspace{-1.5em}
\caption{The relationship between MIMO-based OAC and forward-backward propagation.}
 \vspace{-1.5em}
\label{fig::relationship}
\end{figure*}
Here, we consider a fully connected layer between two nodes implemented by OAC. Note that the bias in a fully connected layer is irrelevant to the input $\bm{x}\in\mathcal{C}^{N_i \times 1}$, hence it can be easily realized on the receiver without extra communication. Therefore, we consider a particular fully connected layer $\bm{y}=\mathbf{W}\bm{x}$ here, where $\mathbf{W}\in\mathcal{C}^{N_o \times N_i}$ and $\bm{y}\in\mathcal{C}^{N_o \times 1}$.

In the MIMO OAC-based system, the input signal $\bm{x}$ is sent through a precoder by the transmitter. Then, the receiver obtains the transmitted signal through a combiner.
Due to channel rank $r$, the number of streams simultaneously transmitted is limited. Hence the receiver can only obtain a $r$-dimentional output signal. By concatenating the output of ${N_o}/r$ transmissions, we can obtain a $N_o$-dimentional output.
Denote the precoder and combiner matrices of the $k$-th transmission by $\mathbf{P}_k\in\mathcal{C}^{N_t \times N_i}$ and $\mathbf{C}^H\in\mathcal{C}_k^{r \times N_r}$, respectively.
We note that the minimum number of transmission rounds ${N_o}/r$ is just a sufficient condition. Since the NN may be sparse and its size may be overlarge, a small number of transmission rounds does not always lead to training failure.

According to the characteristics of MIMO channels, we can easily know that the optimal precoding and combining matrices should follow a specific structure determined by the channel matrix.
In the proposed system, it means that the combining matrix $\mathbf{C}_k$ should be the same, i.e.,  $\mathbf{C}$, since quasi-stable channel is considered. The precoding matrix can also be split into two parts, one for trainable parameters and another for beamforming, i.e., $\mathbf{P}_k=\mathbf{P}\tilde{\mathbf{W}}_k$ where $\mathbf{P}\in\mathcal{C}^{N_t \times r}$ and $\tilde{\mathbf{W}}_k\in\mathcal{C}^{r\times N_i}$.

We then discuss the detailed system implementation and the duality of the reciprocity of wireless MIMO channels and the forward-backward propagation in NNs, as illustrated in Fig. \ref{fig::relationship}.
Hence, the forward computation of the proposed MIMO OAC-based SL can be represented by
\begin{eqnarray}
\bm{x}_{\textrm{t},k}&=&\mathbf{P}\tilde{\mathbf{W}}_k\bm{x},\label{forward1}\\
\bm{y}_{\textrm{r},k}&=&\mathbf{H}\bm{x}_{\textrm{t},k}+\bm{n}_k,\\
\bm{y}_k&=&\mathbf{C}^H\bm{y}_{\textrm{r},k},\label{forward3}
\end{eqnarray}
corresponding to the precoding, transmitting, and combining procedure, respectively, and the output $\bm{y}$ is the concatenation of all $\bm{y}_k$. Ignoring the noise $\bm{n}_k$, we approximately have
\begin{equation}
\bm{y}_k=\mathbf{C}^H\mathbf{H}\mathbf{P}\tilde{\mathbf{W}}_k\bm{x}.\label{4}
\end{equation}

Since the channel $\mathbf{H}$ is of rank $r$,  the overall linear transformation $\mathbf{C}^H\mathbf{H}\mathbf{P}\tilde{\mathbf{W}}_k$ can be equivalent to any parameter $\mathbf{W}_k\in\mathcal{C}^{r\times N_i}$ through adjusting $\tilde{\mathbf{W}}_k$, $\mathbf{P}$, and $\mathbf{C}$.
Therefore, by concatenating the results of $N_o/r$ transmissions, the above scheme can realize any $\mathbf{W}\in\mathcal{C}^{N_o \times N_i}$.

The backpropagation process of the proposed system is introduced as follows.
Based on the basic backpropagation algorithm for complex numbers, the ideal backpropagation corresponding to \eqref{forward1}-\eqref{forward3} should be
\begin{eqnarray}\small
\bm{g}_{C}&=&\sum_{k=1}^K \bm{y}_{\textrm{t},k}\bm{g}_{y_k}^H,\\
\bm{g}_{x_{\textrm{t},k}}&=&\mathbf{H}^H\mathbf{C}\bm{g}_{y_k},\label{back_air}
\end{eqnarray}
and
\begin{eqnarray}\small
\bm{g}_{P_k}&=&\sum_{k=1}^K \tilde{\mathbf{W}}_k\bm{x}\bm{g}_{x_{\textrm{t},k}}^H,\\
\bm{g}_{\tilde{W}_k}&=&\sum_{k=1}^K \bm{x}\mathbf{P}^H\bm{g}_{x_{\textrm{t},k}}^H,\\
\bm{g}_x&=&\sum_{k=1}^K \tilde{\mathbf{W}}_k^H\mathbf{P}^{H}\bm{g}_{x_{t,k}}.
\end{eqnarray}
Using channel reciprocity, the above procedure can be realized by backward OAC without requiring channel $\mathbf{H}$.
If we transmit the conjugation $\bm{g}_{y,k}^*$ by precoding matrix $\mathbf{C}_k^*$ over the backward channel $\mathbf{H}^T$, the received result combining with $\mathbf{P}_k^T$ becomes
\begin{equation}
\tilde{\bm{g}}^*_{x,k}=\mathbf{P}_k^{T}\mathbf{H}^T\mathbf{C}_k^*\bm{g}_{y,k}^H+\mathbf{P}_k^{T}\tilde{\bm{n}}_k.
\end{equation}
Ignoring the noise $\tilde{\bm{n}}_k$ and summarizing the conjugation of the results of all the $K$ backward transmissions, we have
\begin{equation}
\tilde{\bm{g}}_x\approx\left(\sum_{k=1}^K \mathbf{P}_k^{T}\mathbf{H}^T\mathbf{C}_k^*\bm{g}_{y,k}^H\right)^*,\label{backward}
\end{equation}
which is exactly the gradient of the input $\bm{x}$. 
Besides, the gradient of $\mathbf{P}_k$ and $\mathbf{C}_k$ can also be derived in the same way.

The scheme still faces the following two concerns, which can be solved by existing tricks.
Note that there is unavoidable noise in both the forward computation and backpropagation. However, previous work \cite{noisy} shows that noise on the input is tolerable and even becomes a training trick for NNs. Therefore, adding noise during forwarding computation does not seriously affect the learning ability. For backpropagation, introducing noise means using gradients with noise. The well-known stochastic gradient descent algorithm has already shown that using a random gradient with real expectations and limited variance does not deteriorate the performance intensively.
Another concern is about the amplification. In communication systems, most parts may cast amplification on the signal, which means both the forward and backward results are amplified.
In forward computation, batch normalization \cite{BN} layers are adopted, which normalize the result and counteract the amplification.
In backpropagation, we can adopt any optimizer insensitive to the scope of the gradient, such as Adam \cite{adam}.

Before ending this part, we show how to extend the proposed scheme to convolutional NNs, a widely used NN structure, especially in graph processing.
A convolutional layer also casts a linear transformation with a specific structure such as the fully connected layers.
The output of a convolutional layer is typically composed of several channels, and each channel is composed of a 2-dimensional graph with the same size.
Each channel corresponds to a different convolutional kernel.
From the linearity, casting an arbitrary linear transformation to the results of all channels is equivalent to casting the same transformation to all convolutional kernels, which leads to another convolutional layer with the same size.
Hence we can rearrange the intermediate results for transmission such that intermediate outputs corresponding to the same location in each channel are adopted for every transmission, i.e., $\bm{x}$ in \ref{4}.
With such rearrangement, the proposed scheme also works for the general convolutional NNs.
\subsection{Activation by Constellation Diagram}

\begin{figure}[t]
\begin{minipage}[b]{.48\linewidth}
  \centering
  \centerline{\includegraphics[width=4.0cm]{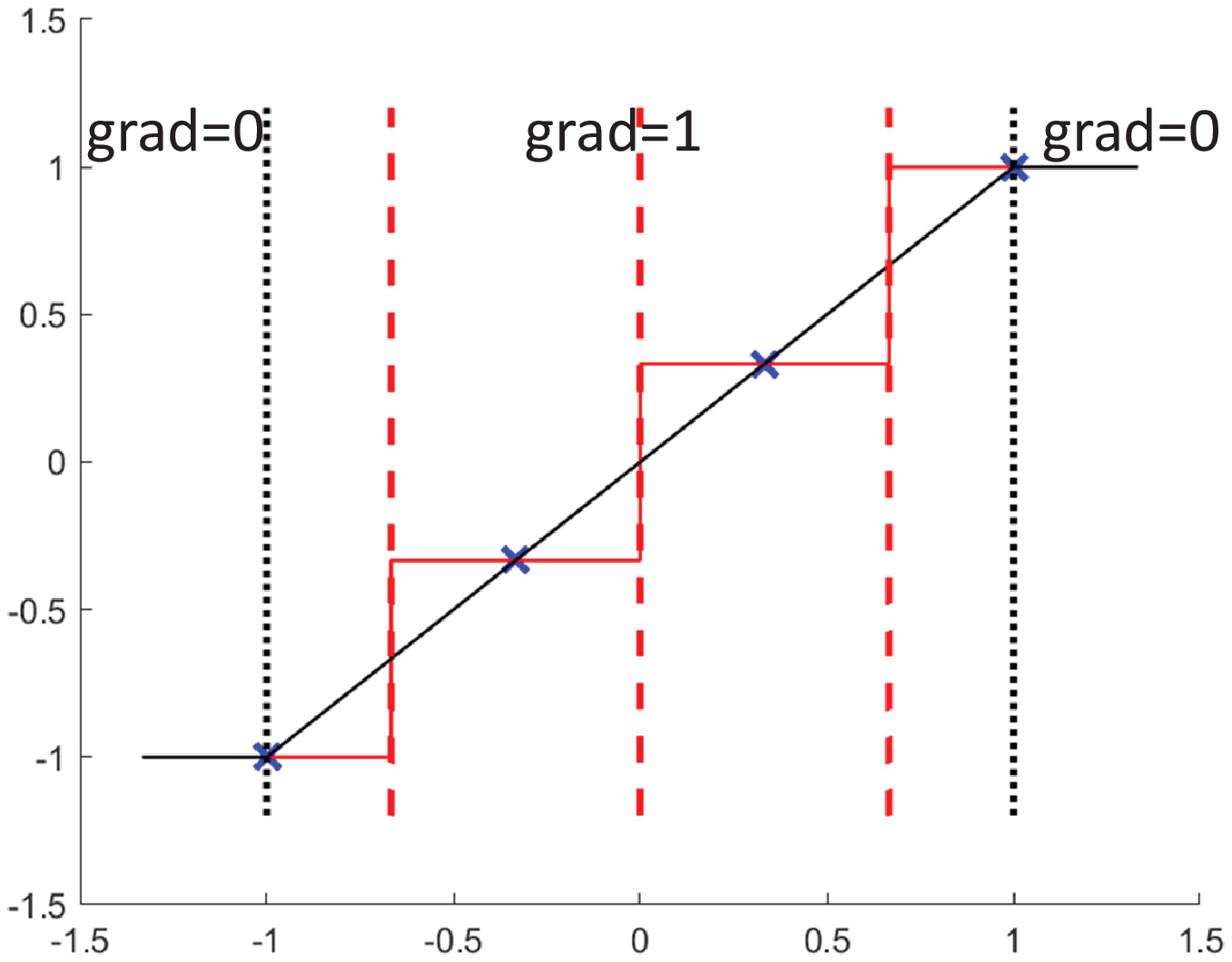}}
  \centerline{(a) 4-level quantification}
  \centerline{for real numbers}\medskip
\end{minipage}
\hfill
\begin{minipage}[b]{0.48\linewidth}
  \centering
  \centerline{\includegraphics[width=4.0cm]{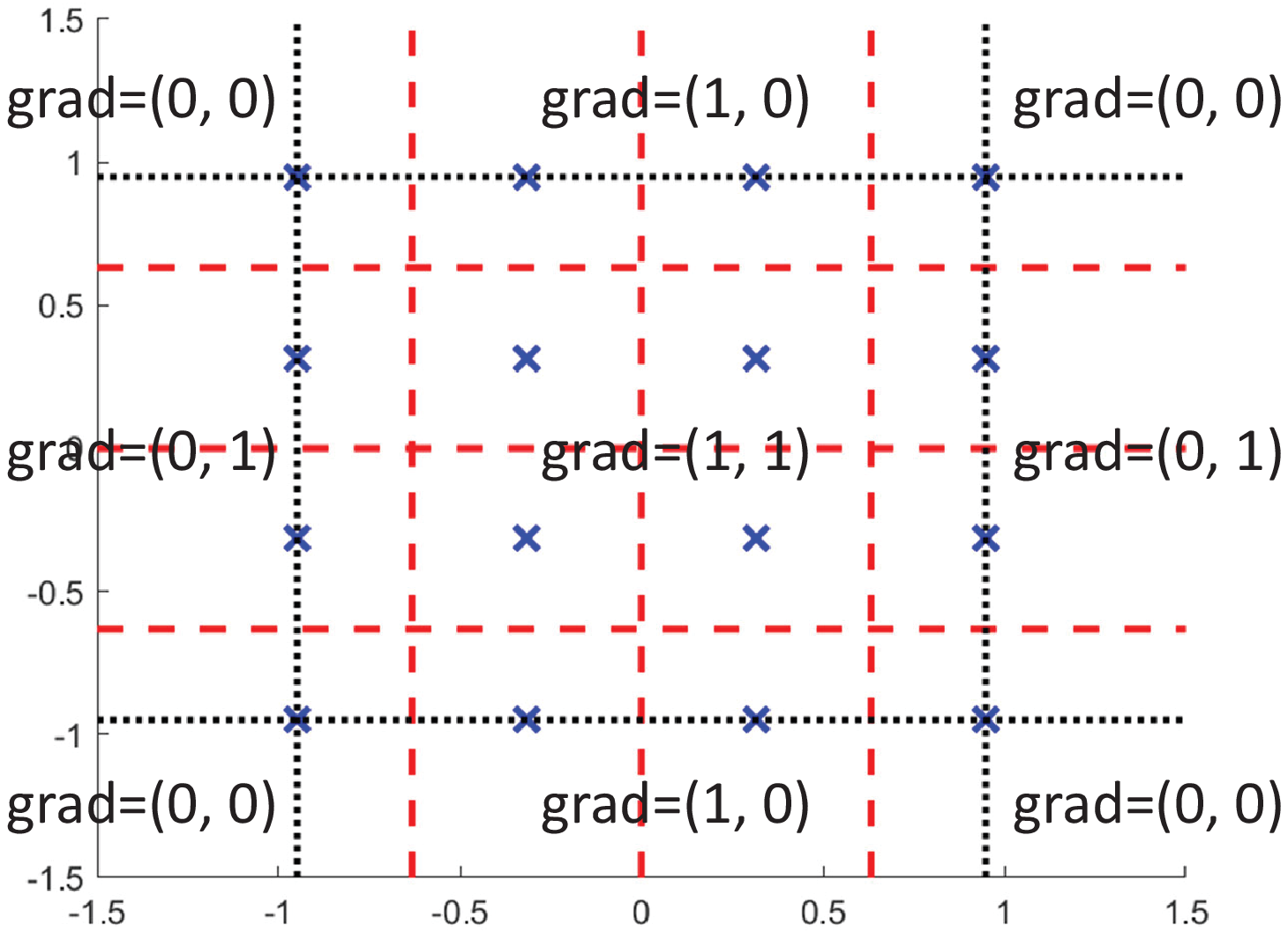}}
  \centerline{(b) 16-QAM quantification}
  \centerline{for complex numbers}\medskip
\end{minipage}
 \vspace{-1.5em}
\caption{The illustration of quantification-based activation functions.}
 \vspace{-1.5em}
\label{fig::activation}
\end{figure}

In wireless communication systems, QAM constellation schemes are used for modulation.
At the receiver side, the obtained signal is quantified to a constellation point before demodulating.
We observe that this quantification by some simple constellation diagrams has been used as some activation functions in quantified NNs as shown in Fig. \ref{fig::activation}.
In binary NNs \cite{BNN}, the activation function can be the step function, which can also be considered as quantification by the binary phase shift keying (BPSK). Through extending the real step function to complex numbers, we get the 4 phase shift keying (4PSK), a widely used modulation scheme in wireless communications. Furthermore, considering multiple-value quantified NNs, there exist other activation functions by multiple-value quantification \cite{DSQ}. Extending these quantification functions to complex numbers, we can obtain QAM constellation diagrams. For example, the 4-value qualification becomes 16QAM. 

Note that in previous works when casting backpropagation through qualification functions, we simply view them as the HardTanh function as shown in Fig. \ref{fig::activation}(a) (the dark line represents  the HardTanh function) for gradient computation.
Fig. \ref{fig::activation}(b) shows the 16QAM-based activation function, where the real and imaginary parts individually adopt the 4-level quantification as in Fig. \ref{fig::activation}(a).
In forward computation, each input is quantified to the nearest constellation point, while the operation differs over the areas divided by the dark lines in backpropagation. 
Mathematically, we can obtain that the $n$-value qualification activation function outputs an element $y\in\{-\Delta,(3-n)\Delta/(n-1),\cdots,\Delta\}$ such that $|y-x|$ takes the minimum value, where $x$ is the input. The corresponding backpropagation is
\begin{equation}
g_x=\left\{\begin{aligned}
0,&\quad y <-\Delta \textrm{ or } y >\Delta\\
g_y,&\quad -\Delta\leq y \leq \Delta
\end{aligned}\right.,
\end{equation}
where $\Delta>0$ is a hyperparameter which usually takes the value of 1.
Extending the procedure to complex numbers, we obtain that for the QAM activation function that outputs the nearest point $y\triangleq y_r +j y_i$ on the constellation diagram for input $x\triangleq x_r +j x_i$, the corresponding backpropagation can be given by
\begin{equation}\small
\mathcal{R}e\{g_x\}=\left\{\begin{aligned}
0,&\quad y_r <-\Delta_r\textrm{ or }y_r >\Delta_r\\
\mathcal{R}e\{g_y\},&\quad -\Delta_r\leq y_r \leq \Delta_r
\end{aligned}\right.,
\end{equation}
\begin{equation}\small
\mathcal{I}m\{g_x\}=\left\{\begin{aligned}
0,&\quad y_i <-\Delta_i\textrm{ or }y_i >\Delta_i\\
\mathcal{I}m\{g_y\},&\quad -\Delta_i\leq y_i \leq \Delta_i
\end{aligned}\right.,
\end{equation}
where $\Delta_r$ and $\Delta_i$ denote the real and imaginary part of the constellation point with the largest power, respectively. From the viewpoint of signal processing, we usually decompose the signal to in-phase and quadrature sinusoidal components. Hence, the above procedure can be easily implemented in the practical systems.

The constellation-based activation functions can be viewed as quantification, which may deteriorate the performance of the NN.
However, it is impractical to generate a signal with an arbitrarily given amplification in the existing communication systems, which is required in previous OAC works.
Through using the constellation-based activation functions, the layers' input is always quantified by a given constellation diagram, which is widely adopted in wireless communication systems.
Note that we use QAM just because it is similar to the multi-level quantification, while other quantification schemes also work.
\section{Training Method}
For the training process, a straightforward way is to directly minimize the loss of the learning task. 
However, this scheme does not take the communication related effect into account. 
Due to the existence of noise, the noise power can affect the training performance especially when the noise power is large.
As a result, a reasonable way is to jointly consider both learning and communication target in the training procedure, i.e., aiming at minimizing the loss function and maximizing  the signal-noise ratio (SNR).

In MIMO communication system, the optimal precoding and combining matrices can be obtained  through calculating singular vectors corresponding to the $r$ largest singular values of the channel $\mathbf{H}$. 
This optimal design also indicates that worst $\min \{N_t,N_r\}-r$ equivalent subchannels should not be used, which are corresponding to the $\min \{N_t,N_r\}-r$ smallest singular values of the channel $\mathbf{H}$.
To avoid using worse subchannels, self-adaptive method is used. 
In the procedure of self-adaptive beamforming, the receiver calculates the covariance matrix $\mathbf{R}_\textrm{f}=\mathbb{E}_k(\bm{y}_k\bm{y}_k^H)$ upon receiving a batch of data, where $\mathbb{E}_k(\cdot)$ stands for the mean value over all transmissions.
Casting single value decomposition (SVD) by $\mathbf{R}_\textrm{f}=\mathbf{U\Sigma V}$, we obtain the loss of forward transmission
\begin{equation}
\ell_{\textrm{f}}=\mathbf{C}\mathbf{U}_{r+1:},
\end{equation}
where $\mathbf{U}_{r+1:}$ represents the $(r+1)$-th to the $N_r$-th columns of $\mathbf{U}$, corresponding to the $N_r-r$  worst equivalent subchannels.
In backward communication, we can similarly obtain the covariance matrix and its SVD as  $\mathbf{R}_\textrm{f}=\mathbb{E}_k(\bm{g}_{x_{\textrm{t},k}}\bm{g}_{x_{\textrm{t},k}}^H)=\mathbf{U'\Sigma' V'}$. The loss is thus
\begin{equation}
\ell_\textrm{b}=\mathbb{E}_k(\bm{x}^H_k\mathbf{V}'_{r+1:}).
\end{equation}
When casting backpropagation, we use the loss $\ell\triangleq\ell_{\textrm{task}}+\ell_{\textrm{f}}+\ell_{\textrm{b}}$ instead of $\ell_{\textrm{task}}$, where $\ell_{\textrm{task}}$ is the loss determined by the learning task.
\section{Numerical Results}\label{sec::num}
We use the complex ResNet \cite{complexResNet} and CIFAR 10 dataset to verify the proposed system.
The complex ResNet consists of an input convolutional layer, six residential blocks with two convolutional layers in each block and a linear output layer.
The NN is split at the 5th and 13th layers. We use batch size 64 and Adam optimizer with a learning rate of 0.005 to train the NN.

An independent 64*64 MIMO system realizes each splitting point.
We adopt the standard MIMO channel model, which is composed of 8 different paths with the gains, transmission, and arrival direction angles independently and randomly generated by uniform distributions, i.e.,
\begin{equation}\label{channel}\small
\mathbf{H} = \sum_{n=1}^{8} a_n (1,\cdots,e^{j(N_r-1)\theta_n})^H(1,\cdots,e^{j(N_t-1)\phi_n}),
\end{equation}
where $\theta_n\sim U(-\pi,\pi)$, $\phi_n\sim U(-\pi,\pi)$, $|a_n|\sim U(0.5, 1.5)$, and $\textrm{angle}(a_n)\sim U(-\pi,\pi)$ for each $n$.
Hence the rank of the channel is approximately $r\approx8$.
\begin{figure}[t]
  \centering
  \centerline{\includegraphics[width=8.5cm]{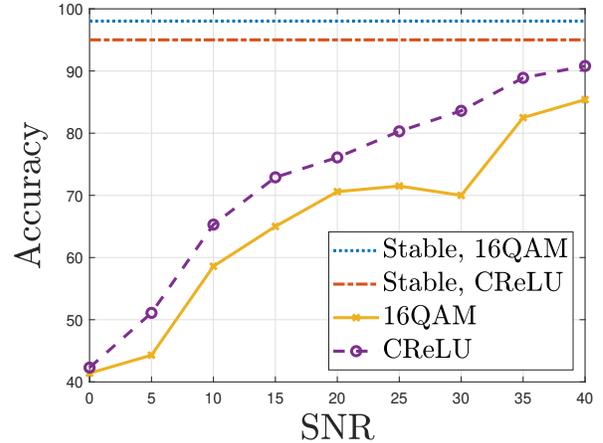}}
%
 \vspace{-1.5em}
\caption{Accuracy performance under various SNRs.}
 \vspace{-1.5em}
\label{fig::num1}
\end{figure}

We consider the proposed system with the traditional complex ReLU activation and the proposed 16 QAM-based activation. To show the effectiveness of the proposed system, we consider the corresponding algorithms where a traditional centralized server implements the NN. We note that the traditional SL systems are equivalent to the centralized ones in terms of performance and require far more communication costs than the proposed system.
\subsection{Results Under Stable Channels}
Here, we discuss the scenario where the channels are stable, are shown in Fig. \ref{fig::num1}.
We can observe that the proposed scheme with both activation functions always performs well.  
The performance gap between the proposed scheme and the corresponding centralized algorithm is marginal for high SNR regime. 
We note that the traditional SL algorithm is strictly equivalent to the centralized algorithm in the aspect of performance. However, the communication cost of the traditional SL algorithm is much higher than the proposed scheme, for example, 
the communication cost of SL algorithm is at least 32 times higher than that in the proposed system since a float number is always represented by 32 bits.
It can be also found that there is always a gap of about 10 \% in the performance of both activation functions.
This is mainly because that 16QAM is a quantification with pretty low precision. Therefore, it cannot perform as well as traditional activation functions.
However, by reasonably designing the NN structure as that in quantification NNs, the performance of NN with 16QAM can be improved \cite{BNN, DSQ}.
\subsection{Results Under Varying Channels}
We consider a mobile channel
\begin{equation}
\mathbf{H}_{t+1}=(1-\rho)\mathbf{H}_t + \rho \tilde{\mathbf{H}}_t,\label{moving}
\end{equation}
where $\tilde{\mathbf{H}}_t$ is independently and randomly generated by \eqref{channel}, and $\rho$ is a constant parameter to indicate the rapidness of the channel varying.
The channel varies using \eqref{moving} every 50 training batches.
With different values of $\rho$, the channel model \eqref{channel} represents varying channels with different speeds.
\begin{figure}[t]
  \centering
  \centerline{\includegraphics[width=8.5cm]{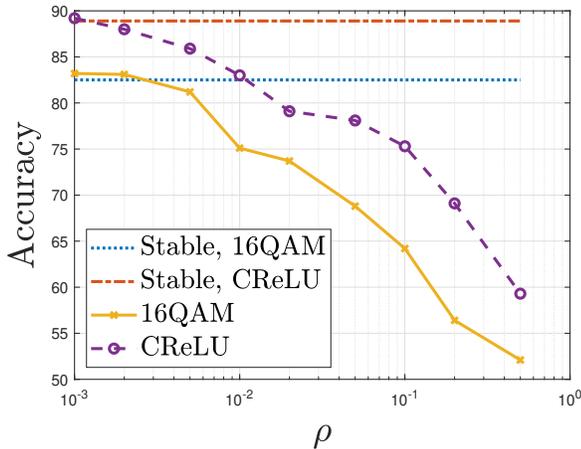}}
%
 \vspace{-1.5em}
\caption{Accuracy performance versus the value of mobile channel parameter $\rho$.
}
 \vspace{-1.5em}
\label{fig::num2}
\end{figure}

In Fig. \ref{fig::num2}, we show the best results over the first 50 epochs of the proposed scheme with both CReLU and 16QAM activation functions.
It is  shown that the proposed scheme with both activation functions works well when the channel varies slowly.
\vspace{-.5cm}
\section{Conclusion}
In this paper, we proposed an OAC-based SL scheme with MIMO-based NNs and activation functions based on constellation diagrams, which integrates communication and computation and significantly improves the overall efficiency. 
In the proposed system, the MIMO wireless channel with adaptive precoding and combining matrices can be viewed as a fully connected layer in NNs.
We proved the interaction between the channel reciprocity in wireless communication and 
forward-backward propagation in NNs. Hence, the channel estimation can be omitted in the proposed system, greatly reducing the system cost.
Moreover, we revealed that constellation diagrams are native 
activation functions in NNs.
Numerical results show the effectiveness of the proposed scheme. 

There remain several future directions for this work.
Firstly, we only considered strict channel reciprocity in this paper and the extension to natural channels with only path reciprocity remains a future direction.
Moreover, by some semantic information theory-based algorithms, the effect of noise may be declined, which can be combined with the proposed scheme.
We only consider the quantification in the forward computation in this paper. As for the backpropagation, quantification may also be applied without seriously degrading the performance since the gradients just indicate directions of optimization, which do not require very high precision.
The proposed scheme can also be combined with network for AI applications
such as distributed sensing, channel estimation and prediction, location estimation and prediction, distributed network optimization, interference coordination, etc.
\vspace{-0.5cm}
\bibliographystyle{IEEEbib}
\bibliography{strings,refs}

\end{document}